\documentstyle[12pt]{article}

\newcommand{\bce}{\begin{center}}
\newcommand{\ece}{\end{center}}
\newcommand{\beq}{\begin{equation}}
\newcommand{\eeq}{\end{equation}}
\newcommand{\bea}{\vspace{0.25cm}\begin{eqnarray}}
\newcommand{\eea}{\end{eqnarray}}

\newcommand{\ba}{\begin{array}}
\newcommand{\ea}{\end{array}}

\newcommand{\doublespace}{
    \renewcommand{\baselinestretch}{1.6}\large\normalsize}

\def\lsim{\mathrel{\rlap{\lower4pt\hbox{\hskip1pt$\sim$}}
    \raise1pt\hbox{$<$}}}         %less than or approx. symbol
\def\gsim{\mathrel{\rlap{\lower4pt\hbox{\hskip1pt$\sim$}}
    \raise1pt\hbox{$>$}}}         %greater than or approx. symbol

\def\Pom{{\bf I\!P}}

\def\beq{\begin{equation}}
\def\endeq{\end{equation}}
\def\arr{\begin{eqnarray}}
\def\endarr{\end{eqnarray}}
\makeindex

  \advance\oddsidemargin  by -1in
  \advance\evensidemargin by -1in
  \advance\topmargin      by -0.5in
\begin{document}

\phantom{.}{\large \bf \hspace{8.5cm} KFA-IKP(Th)-1994-37 \\
\phantom{.}\bf \hspace{10.2cm} DFTT 42/94 \\
\phantom{.}\hspace{9.9cm}5 October 1994\vspace{.4cm}\\
}

\begin{center}
{\Large \bf Diffractive DIS from the
generalized BFKL pomeron. Predictions for HERA.
\vspace{1.0cm}\\}
{\large \bf M.Genovese$^{a}$, N.N.~Nikolaev$^{b,c}$,
B.G.~Zakharov$^{b,c}$ \vspace{1.0cm}\\}
{\it
$^{a}$ Dipartimento di Fisica Teorica, Universit\`a di Torino,\\
and INFN, Sezione di Torino, Via P.Giuria 1, I-10125 Torino, Italy
\medskip\\
$^{b}$IKP(Theorie), KFA J{\"u}lich, 5170 J{\"u}lich, Germany
\medskip\\
$^{c}$L. D. Landau Institute for Theoretical Physics, GSP-1,
117940, \\
ul. Kosygina 2, Moscow V-334, Russia.\vspace{1.0cm}\\ }
{\Large \bf
Abstract}\\
\end{center}
We present microscopic QCD calculation of the cross
section of diffractive DIS and of the partonic structure of
the pomeron from the dipole approach to the generalized BFKL
pomeron. We show that the pomeron can not be treated as a
particle with uniquely defined structure function and
flux in the proton. We find strong factorization breaking
which can approximately be described by the two-component
structure function of the pomeron, each component
endowed with the different flux of pomerons in the proton.
We predict very weak $Q^{2}$ dependence of the
diffractive contribution to the proton
structure function.
\bigskip\\

\begin{center}
E-mail: kph154@zam001.zam.kfa-juelich.de
\end{center}

 \doublespace
\pagebreak

%---------------------------------------------------

\section{Introduction}

%-------------------------------------------
%                Section 1
%-------------
Much progress
in our understanding of the QCD pomeron is expected
from experiments on diffractive deep inelastic scattering (DIS) in
progress at HERA. A detailed description of diffractive DIS
in terms of the diffraction excitation of multiparton Fock
states of the photon,
\beq
\gamma^{*}+p\rightarrow X+p'\, .
\label{eq:1.1}
\endeq
which interact with the target proton by the
dipole BFKL pomeron exchange [1-5], was developed in [1-3].
Extrapolating the Regge theory considerations [6], one can
alternatively view inclusive reaction (\ref{eq:1.1}) as DIS
on pomerons radiated by protons. This analogy inspired suggestions
[7,8], although conspicuously short of the microscopic QCD
derivation, to treat pomeron as a particle with a well
defined partonic structure. Understanding the accuracy, and
limitations, of such a partonic
description of inclusive diffractive
DIS is a topical issue which we address here in the framework
of the microscopic dipole-cross section approach to
the generalized BFKL pomeron [1-5]. Our
principal conclusion is that this
only is possible
at the expense of a
two-component partonic structure of the pomeron, which
leads to a specific breaking of the conventional
parton-model factorization.

We consider DIS
at $x={Q^{2}\over (Q^{2}+W^{2})} \ll 1$, followed by
diffraction
excitation of the virtual photon into the state $X$ of mass $M$,
where $Q^{2}$ is the virtuality of the photon and $W$ is the total
energy in the photon-proton center of mass.
The
variable $x_{\Pom}={(M^{2}+Q^{2}) \over (W^{2}+Q^{2})}\ll 1$
can be interpreted as
a fraction of proton's momentum taken away by the pomeron, whereas
$\beta = {Q^{2}\over (Q^{2}+M^{2})}$
is the Bjorken variable for DIS on
the pomeron. Notice that
\beq
x_{\Pom}\beta = x  \, .
\label{eq:1.2}
\endeq
The final-state proton
$p'$ carries the fraction $(1-x_{\Pom})$ of the beam proton's momentum
and is separated from the hadronic debris $X$ of the photon
by the (pseudo)rapidity gap $\Delta \eta \approx \log{1\over x_{\Pom}}
\geq \Delta \eta_{c}  \gsim $(2.5-3).
In the following, we take $x_{\Pom}=x_{\Pom}^{0}=0.03$ as
reference point.
Once the total cross section of photoabsorption on the pomeron
$\sigma_{tot}(\gamma^{*}\Pom,M^{2})$ is known, the pomeron structure
function can operationally be defined by the standard formula
\beq
F_{2}^{(\Pom)}(x,Q^{2})={Q^{2}\over 4\pi^{2}\alpha_{em}}
\sigma_{tot}(\gamma^{*}\Pom,M^{2})\, .
\label{eq:1.3}
\endeq
The experimentally measured quantity is
$d \sigma_{D}(\gamma^{*}\rightarrow X)
/dt dM^{2}$, where $t$ is the $(p,p')$ momentum transfer squared.
Under the assumption of single-pomeron exchange, generalization
of the Regge theory convention [6] gives the operational definition
[2,3]
\beq
\sigma_{tot}(\gamma^{*}\Pom,M^{2})=
{16 \pi \over \sigma_{tot}(pp)}( M^{2} +Q^{2})
\left. { d\sigma_{D}(\gamma^{*}+p \rightarrow X+p)
 \over dt dM^{2}}\right|_{t=0} \, .
\label{eq:1.4}
\endeq
This convention
assumes that total cross section is asymptotically
constant, {\sl i.e.}, the flux of pomerons in the proton
$f_{\Pom}(x_{\Pom})/x_{\Pom}$
satisfies $f_{\Pom}(x_{\Pom})=1$. The generalization of
(\ref{eq:1.4}) to DIS, under the {\sl strong assumption} of
factorization of the flux and structure function of pomerons is
\footnote{Ref.~[1] used the factor $M^{2}$ instead
of the $(M^{2}+Q^{2})$ in the {\sl l.h.s.} of Eq.~(\ref{eq:1.4}).
Other conventions are possible [7,8], but the observable cross
sections do not depend on how one factorizes them into the flux
and structure function of pomerons.}
\arr
\left.(M^{2}+Q^{2})
{ d\sigma_{D} (\gamma^{*}\rightarrow X)
\over dt\,d M^{2} }\right|_{t=0} =
\left.x_{\Pom}{ d\sigma_{D} (\gamma^{*}\rightarrow X)
\over dt\,dx_{\Pom} }\right|_{t=0} =
{  \sigma_{tot}(pp) \over 16\pi}
{4\pi^{2} \alpha_{em}
\over Q^{2}}
f_{\Pom}(x_{\Pom}) F_{2}^{(\Pom)}(\beta,Q^{2})\, ,
\label{eq:1.5}
\endarr
where $\alpha_{em}$ is the fine structure constant,
$\sigma_{tot}(pp)=40$\,mb is an energy-independent
normalization constant and hereafter we
make use of relation
(\ref{eq:1.2}). Evidently, the above set of {\sl
operational
definitions} only makes sense if the pomeron
flux function $f_{\Pom}(x_{\Pom})$ can be defined
in such a way that
the $Q^{2}$ dependence of the {\sl r.h.s.} of (\ref{eq:1.5})
is concentrated in
$F_{2}^{\Pom}(\beta,Q^{2})$, which
satisfies the conventional QCD evolution.
This factorization (convolution) property (\ref{eq:1.5}) and
the  QCD evolution
property of $F_{2}^{\Pom}(\beta,Q^{2})$
must be proven
starting with the microscopic QCD treatment of diffractive DIS
rather than be postulated.

The QCD (BFKL [9]) pomeron is described by the (generalized)
BFKL equation which recently was reformulated in the dipole-cross
section representation [2-5] (somewhat related approach is also
discussed in [10]). In this dipole BFKL approach, the convolution
representation (\ref{eq:1.5}) is problematical for many reasons.
For instance, at subasymptotic energies, the dipole pomeron
does not factorize [1-3], and
the recent BFKL phenomenology of
DIS has shown [11] that the kinematical domain of HERA is
the subasymptotic one. Furthermore, the naive partonic
description of the pomeron was shown to fail in the diffractive
jet production [1,12]. In this communication we demonstrate
that, indeed, the convolution (\ref{eq:1.5}) breaks down,
but a sort of factorization is restored in a two-component
picture, in which the pomeron is endowed with
two structure
functions, which evolve, according to GLDAP equations [13],
from the initial valence quark-antiquark
and the valence gluon-gluon components of the pomeron,
respectively. For these
two components, the fluxes of pomerons in the proton are different.
In striking contrast to the case
of hadrons, normalizations of the glue and
sea components of the pomeron contain the dimensionfull
(triple-pomeron) coupling $A_{3\Pom}^{*}$ which has the dimension
$[{\rm GeV}]^{-2}$ and absorbs the ensuing infrared sensitivity in
the problem. (For the direct
calculation of $A_{3\Pom}^{*}$, discussion of its relation to
the conventional triple-pomeron coupling $A_{3\Pom}(Q^{2})$ and
a comparison of $A_{3\Pom}(Q^{2})$ in DIS and real photoproduction
see [14]).

The further presentation is organized as follows: In section 2
we derive the valence $q\bar{q}$ structure function of the
pomeron and the corresponding flux of pomerons
$\phi_{\Pom}(x_{\Pom})$
in the proton. In section 3 we derive the sea structure function
of the pomeron and the corresponding flux
$f_{\Pom}(x_{\Pom})$, which is different
from the $\phi_{\Pom}(x_{\Pom})$.
In section 4 we formulate the two-component
description of the pomeron structure function and discuss
the breaking of factorization (\ref{eq:1.5}). Predictions
for the diffractive contribution $F_{2}^{D}(x,Q^{2})$ to
the proton structure function are presented in Section 5.
In the Conclusions section we summarize our major results.
%-------------------------------------
\section{The
valence quark-antiquark component of the pomeron}

%-------------------------------------
%                      Section 2
%---------

The approach [1-3] starts with the microscopic QCD calculation
of $d\sigma_{D}/dt dM^{2}|_{t=0}$ and a thorough examination
of whether
it can be reinterpreted, via Eqs.~(\ref{eq:1.3}-\ref{eq:1.5}),
in terms of a GLDAP evolving
pomeron structure function or not. Diffraction
excitation of the $q\bar{q}$ Fock state of the photon (Fig.1a)
has the cross section (hereafter we
focus on the dominant diffraction dissociation of transverse
photons)
\arr
\left.{d\sigma_{D}(\gamma^{*}\rightarrow X)\over dt}
\right|_{t=0}=
\int dM^{2}\,
\left.{d\sigma_{D}(\gamma^{*}\rightarrow X)\over dtdM^{2}}
\right|_{t=0}~~~~~~~~~~~~~~~~~~~~~~~\nonumber \\
= {1 \over 16\pi}
\int_{0}^{1} dz\int d^{2}\vec{r}\,\,
\vert\Psi_{\gamma^{*}}(Q^{2},z,r)\vert^{2}\sigma(x,r)^{2}
= {4\pi^{2} \alpha_{em} \over  Q^{2}}
\int {dr^{2}\over r^{2}} W(Q,r)
\left[{\sigma(x,r)\over r^{2}}\right]^{2}
\,\,\,.
\label{eq:2.1}
\endarr
Here $\vec{r}$ is the transverse separation of the quark and
antiquark in the photon, $z$ and $(1-z)$ are partitions of
photon's lightcone momentum between the quark and antiquark,
$\sigma(x,r)$ is the dipole cross section for scattering on
the proton target (hereafter we use $\sigma(x,r)$ of
Refs.~[11,15]), and the dipole distribution in the photon
$\vert\Psi_{\gamma^{*}}(Q^{2},z,r)\vert^{2}$ derived in [16], equals
\beq
\vert\Psi_{\gamma^{*}}(Q^{2},z,r)\vert^{2}
={6\alpha_{em} \over (2\pi)^{2}}
\sum_{i}^{N_{f}}e_{i}^{2}
\{[z^{2}+(1-z)^{2}]\varepsilon^{2}K_{1}(\varepsilon r)^{2}+
m_{q}^{2}K_{0}(\varepsilon r)^{2}\}\,\,,
\label{eq:2.2}
\endeq
where $\alpha_{em}$ is the fine structure constant, $e_{i}$ is the
quark charge in units of the electron charge, $m_{q}$ is the quark
mass, $\varepsilon^{2} = z(1-z)Q^{2}+m_{q}^{2}$ and $K_{\nu}(x)$
is the modified Bessel function. Precisely the
same dipole cross section enters the calculation of the proton
structure function
\beq
F_{2}^{p}(x,Q^{2})= {Q^{2} \over 4\pi^{2}\alpha_{em}}
\int_{0}^{1} dz\int d^{2}\vec{r}\,\,
\vert\Psi_{\gamma^{*}}(Q^{2},z,r)\vert^{2}\sigma(x,r)\,\,,
\label{eq:2.3}
\endeq
and the scenario [11,15] for $\sigma(x,r)$ was shown [11] to give
a good quantitative description of the HERA data [17].

The ingredients which allow reinterpretation of the cross section
(\ref{eq:2.1}) as DIS on the valence $q\bar{q}$ state of the pomeron
are: \\
(i) The mass spectrum calculated in [1], which roughly follows
\beq
\left.{d\sigma_{D}\over dM^{2}dt}\right|_{t=0}
\propto {M^{2}\over (Q^{2}+M^{2})^{3}  }=
{1\over Q^{4}}\beta^{2}(1-\beta)
\, ,
\label{eq:2.4}
\endeq
is to a good approximation $x$-independent. \\
(ii) At large $Q^{2}$ the weight function
$W(Q,r)$ is $Q^{2}$-independent
and the diffractive cross section $\sigma_{D}(\gamma^{*} \rightarrow
q\bar{q})$ satisfies the Bjorken scaling [1,16].\\
(iii) The weight function $W(Q,r)$ is peaked at large,
and $Q^{2}$-independent, hadronic size
 $r=R_{val} \sim 1/m_{q}$. There is much semblance to the
$Q^{2}$-independent spatial separation of the valence quark and
antiquark in the pion, and
we can analogously speak of DIS off the valence $q\bar{q}$ state
of the pomeron.
The corollary is
that in
Eq.~(\ref{eq:2.1}) the $x$ and $\beta$ dependence
can be factorized and
we can write down the convolution representation
\beq
\left.x_{\Pom}
{ d\sigma_{D}(\gamma^{*}\rightarrow q\bar{q})
 \over dt\,d x_{\Pom} }\right|_{t=0} =
{  \sigma_{tot}(pp) \over 16\pi}\cdot
{4\pi^{2} \alpha_{em}
\over Q^{2}}
\phi_{\Pom}(x_{\Pom})
F_{val}^{\Pom}(\beta)\,\, ,
\label{eq:2.5}
\endeq
in which the valence $q\bar{q}$ structure function of the pomeron
\beq
F_{val}^{(\Pom)}(\beta)=C_{val}\beta(1-\beta)=0.27\beta(1-\beta)
\label{eq:2.6}
\endeq
follows from the mass spectrum (\ref{eq:2.4}) ([1], see also [8]),
and the flux function $\phi_{\Pom}(x_{\Pom})$ is defined by
\beq
\phi_{\Pom}(x_{\Pom})={
\int_{0}^{1} dz\int d^{2}\vec{r}\,\,
\vert\Psi_{\gamma^{*}}(Q_{\Pom}^{2},z,r)\vert^{2}
\sigma(x_{\Pom},r)^{2}
\over
\int_{0}^{1} dz\int d^{2}\vec{r}\,\,
\vert\Psi_{\gamma^{*}}(Q_{\Pom}^{2},z,r)\vert^{2}
\sigma(x_{\Pom}^{0},r)^{2}
} \,
\label{eq:2.7}
\endeq
subject to the normalization $\phi_{\Pom}(x_{\Pom}^{0}=0.03)=1$.
For the definition of the factorization scale $Q_{\Pom}^{2}$ see
below; at large $Q_{\Pom}^{2}$ the flux function $\phi(x_{\Pom})$
does not depend on $Q^{2}_{\Pom}$.
Because in DIS on the valence (anti)quarks $\beta \sim 1$, see
Eq.~(\ref{eq:2.6}), at $x\ll 1$ we can neglect
the distinction between $\phi_{\Pom}(x)$ and $\phi_{\Pom}(
x_{\Pom}={x\over\beta})$ in (\ref{eq:2.5}).

Both the normalization $C_{val}$ of $F_{val}^{\Pom}(\beta)$
and the $x_{\Pom}$-dependence of the pomeron
flux function $\phi_{\Pom}(x_{\Pom})$ are controlled by
$\sigma(x,r)$ at the
large, nonperturbative, dipole size $r\sim {1\over m_{q}}$.
We take $m_{q}=0.15$\,GeV, which gives a good quantitative
description of the real photoabsorption cross section [15] and
nuclear shadowing in DIS [16,18], which are controlled by
a similar dipole size.
The flux function $\phi_{\Pom}(x_{\Pom})$ is
shown in Fig.~2. The absolute normalization of the valence of
the pomeron,
$C_{val}= 0.27$,
in (\ref{eq:2.6}) is fixed requiring
that the convolution (\ref{eq:2.5}) gives the same $q\bar{q}$
excitation cross section as formula (\ref{eq:2.1}).
The flavour decomposition of valence parton distributions
${\rm v}_{i}(\beta)=A_{i}(1-\beta)$ is $A_{u}=A_{\bar{u}}=A_{d}=
A_{\bar{d}}= 0.20$, $A_{s}=A_{\bar{s}}=0.11$,
$A_{c}=A_{\bar{c}}=0.02$ (for a discussion of the flavour
asymmetry of diffractive DIS see [1])
\footnote{Because of a different convention, see footnote~1,
the structure function $F_{val}^{\Pom}(NZ92,\beta)
=0.25\beta(1-\beta)^{2}$ of Eq.~(50) in Ref.~[1] contains the
extra factor $(1-\beta)$, apart from that $F_{val}^{\Pom}(NZ92,\beta)$
is identical to $F_{val}^{\Pom}(\beta)$ of the
present paper.}. A conservative estimate of the
uncertainty in $\sigma(x_{\Pom},r\sim {1\over m_{q}})$ is
$\lsim$(15-20)\%, and the uncertainty in our prediction for
$C_{val}$ is $\lsim$30\%. These valence distributions
can be used as an  input at $Q^{2}=Q_{\Pom}^{2} = 10$\,GeV$^{2}$
(this choice is discussed below)  for the GLDAP evolution of
$F_{val}^{\Pom}(\beta,Q^{2})$, which sums the higher order
diagrams of Fig.~1b, describing the sea
originating from the pure valence $q\bar{q}$
pomeron. The predicted
$Q^{2}$-dependence of $F_{val}^{\Pom}(\beta,Q^{2})$
is shown in Fig.~3.

\section{Valence gluons and sea in the pomeron}

%----------------------------------------------
%                            Section 3
%------------
The mass spectrum (\ref{eq:2.4}) for excitation of the $q\bar{q}$
state rapidly decreases at large $M^{2} \gg Q^{2}$. The $\sim
1/M^{2}$ mass spectrum, typical of the so-called triple-pomeron
regime [8,19], first emerges from diffractive excitation of the
$q\bar{q}g$ Fock state of the photon in Fig.~1c [1-3]. The new
parameter which emerges in the lightcone description of the
$q\bar{q}g_{1}...g_{n}$ Fock states of the photon is the
correlation (propagation) radius $R_{c}=1/\mu_{G}$ for the
perturbative gluons. Following [5,11,12], we take $R_{c}\approx 0.27
$\,fm ($\mu_{G}=0.75$\,GeV) as suggested by lattice QCD studies [20].
The major finding of [2,3] is that, at large
$Q^{2} \gg 1/R_{c}^{2}$, the cross section of diffraction excitation
$\gamma^{*}\rightarrow q\bar{q}g$ takes on the form
%marco
(where $\alpha_{S}(r)$ is the running QCD coupling in function of r
[2])
\arr
(Q^{2}+M^{2})\left.{d\sigma_{D} \over dt dM^{2}}\right|_{t=0}
\simeq \int dz \,d^{2}\vec{r}\,\,
|\Psi_{\gamma^{*}}(Q^{2},z,r)|^{2}\cdot {16\pi^{2} \over 27}\cdot
\alpha_{S}(r) r^{2} \nonumber\\
\times {1\over 2\pi^{4}}\cdot\left({9\over 8}\right)^{3}
\cdot \int d\rho^{2}
\left[{\sigma(x_{\Pom},\rho)\over \rho^{2}}\right]^{2}
{\cal F}(\mu_{G}\rho)
\label{eq:3.3}
\endarr
with factorized $Q^{2}$ and $x_{\Pom}$ dependence.
In (\ref{eq:3.3}), the form factor
${\cal F}(z)=
z^2 \cdot [ K_{1}(z)^2+z K_{1}(z)K_{0}(z)+{1 \over 2} z^2 K_{0}(z)^2]$.
It is precisely this factorization property which allows to
define the corresponding structure function and the pomeron
flux function.

It is convenient to introduce
the normalization constant $A_{3\Pom}^{*}$ such that
\beq
A_{3\Pom}^{*}f_{\Pom}(x_{\Pom})= {1\over 2\pi^{4}}\cdot
\left({9\over 8}\right)^{3} \cdot \int dr^{2}
\left[{\sigma(x_{\Pom},r)\over r^{2}}\right]^{2}
{\cal F}(\mu_{G}r) \, ,
\label{eq:3.4}
\endeq
where $f_{\Pom}(x_{\Pom})$ is the corresponding
flux function, subject to the
normalization $f_{\Pom}(x_{\Pom}^{0})=1$. The constant
$A_{3\Pom}^{*}=0.56$\,GeV$^{-2}$ has a meaning, and the
magnitude close to that, of the triple pomeron coupling
$A_{3\Pom}(Q^{2})$ (for the more detailed discussion see [14]).
Furthermore, one can introduce the explicit two-gluon wave
function of the pomeron [2,3]
\beq
|\Psi_{\Pom}(\beta,\vec{r})|^{2} = {1\over f_{\Pom}(x_{\Pom})}\cdot
{81 \over 8 \pi^{4}}
\cdot {1-\beta \over \beta}\cdot{1\over \sigma_{tot}(pp)}
\left[{\sigma(x_{\Pom},r)\over r^{2}}\right]^{2}
{\cal F}(\mu_{G}r) \, ,
\label{eq:3.1}
\endeq
where $\vec{r}$ is the transverse separation of gluons in the
pomeron and $\beta$ is a fraction of pomeron's momentum carried
by a gluon. In the wave function
(\ref{eq:3.1}), the
$x_{\Pom}$-dependence cancels out approximately, and it gives
the $x_{\Pom}$-independent
gluon structure function of the pomeron
\beq
G_{\Pom}(\beta)=\beta g_{\Pom}(\beta)=
\beta \int d^2\vec{r}\, |\Psi_{\Pom}(\beta,\vec{r})|^{2} =
A_{G}(1-\beta)\, .
\label{eq:3.2}
\endeq
The wave function
(\ref{eq:3.1}) corresponds to a relatively small
transverse size of the $gg$ state of the pomeron $r\sim R_{sea}
\approx R_{c}$. In DIS on protons, the onset of GLDAP evolution
requires
$Q^{2}\gsim Q_{N}^{2}\sim 2$GeV$^{2}$ [21]. Then
we can argue that, in DIS on
the pomeron, GLDAP evolution becomes applicable at $Q^{2}
\gsim Q_{\Pom}^{2} =Q_{N}^{2}({R_{p}/R_{sea}})^{2}$. As
factorization scale for the pomeron, we take $Q_{\Pom}^{2}=
10$\,GeV$^{2}$. Then,
Eq.~(\ref{eq:3.3})
gives the input sea structure function of the pomeron
\arr
F_{sea}^{\Pom}(\beta \ll 1, Q_{\Pom}^{2})=
C_{sea}=
{16\pi A_{3\Pom}^{*} \over \sigma_{tot}(pp)}\cdot
{Q_{\Pom}^{2} \over 4\pi^{2}\alpha_{em}}\cdot
\int_{0}^{1}dz d^{2}\vec{r} \,
|\Psi_{\gamma^{*}}(Q_{\Pom}^{2},r,z)|^{2}
\cdot{16\pi^{2} r^{2}\alpha_{S}(r) \over 27} \, .
\label{eq:3.5}
\endarr
%marco where $\alpha_{S}(r)$ is the running QCD coupling.
Following
[2,3], one can easily show that
\beq
F_{sea}^{\Pom}(\beta,Q_{\Pom}^{2}) \propto
\log\left[{1 \over \alpha_{S}(Q^{2}_{\Pom})}\right]  \, ,
\label{eq:3.6}
\endeq
which is the correct QCD scaling violation for the sea structure
function which evolves from the valence gluonic state.

In
(\ref{eq:3.1}), the $\propto 1/\beta$ dependence is a rigorous
result [2,3],
the factor $(1-\beta)$ in
(\ref{eq:3.1}) is
an educated guess. At $\beta \rightarrow 1$ it makes,
in the spirit of the
quark counting rules,
the two valence distributions (\ref{eq:2.6}) and (\ref{eq:3.1})
similarly behaved, and $q\bar{q}$ sea contribution to the
pomeron structure function behaved as $\sim (1-\beta)^{2}$.
As a starting approximation, we take
\beq
F_{sea}^{\Pom}(\beta,Q_{\Pom}^{2})=C_{sea}(1-\beta)^{2}=
0.063(1-\beta)^{2}
\label{eq:3.7}
\endeq
with the normalization which follows from
Eq.~(\ref{eq:3.5}) and only slightly differs from
the estimate of Ref.~[1].
The flavour decomposition of the input sea in the pomeron is
$q_{sea}^{(i)}(\beta,Q_{\Pom}^{2})=
\bar{q}_{sea}^{(i)}(\beta,Q_{\Pom}^{2})= A_{sea}^{(i)}(1-\beta)^{2}$,
is $A_{sea}^{(u)}=A_{sea}^{(d)}=0.048,~A_{sea}^{(s)}=0.040,
{}~A_{sea}^{(c)}=0.009$. Finally,
\beq
A_{G}= \int d^2\vec{r}
\left\{\beta |\Psi_{\Pom}(\beta,\vec{r})|^{2}\right\}_{\beta=0}
 =
{128 \pi \over 9}\cdot {A_{3\Pom}^{*} \over \sigma_{tot}(pp) } =
0.28 \, .
\label{eq:3.9}
\endeq
This fully specifies the (parameter-free) input for the
$Q^{2}$ evolution of the pomeron structure function $F_{sea}^{\Pom}
(\beta,Q^{2})$, which originates from the gluonic component
of the pomeron. QCD evolution sums the diagrams of Fig.~1(c,d),
the result of evolution of $F_{sea}^{\Pom}(\beta,Q^{2})$
is shown in Fig.~3. $F_{sea}^{\Pom}(\beta,Q^{2})$ takes over
$F_{val}^{\Pom}(\beta,Q^{2})$ at $\beta \lsim 0.2$, as it was
predicted in [1].

Regarding the accuracy of our estimates for $C_{sea}$ and $A_{G}$,
we wish to recall that real photoproduction of the $J/\Psi$ [22],
exclusive leptoproduction of the $\rho^{0}$ at $Q^{2}\lsim
$(10-20)GeV$^{2}$ [23] and color transparency effects in the
$J/\Psi$ [24] and $\rho^{0}$ [25] production on nuclei probe the
(predominantly nonperturbative) dipole cross section at $r\sim
0.5$\,fm$\lsim 2R_{c}$ [15,26-28]. Real, and weakly virtual $Q^{2}
\sim 10$\,GeV$^{2}$, photoproduction of the open
charm probes the (predominantly perturbative) dipole
cross section at $r\sim {1\over m_{c}}\sim {1\over 2}R_{c}$ [11,15].
The proton structure function $F_{2}^{p}(x,Q^{2})$ probes the
dipole cross section in a broad range of radii from $r\sim 1$\,fm
down to $r\sim 0.02$\,fm [11,21]. Successful quantitative
description of the corresponding experimental data in [11,15,26-28]
implies that we know the dipole cross section
$\sigma(x_{\Pom},r\sim R_{c})$ to a conservative uncertainty
$\lsim $(15-20)\%, and our predictions for $C_{sea}$ and $A_{G}$
have an accuracy $\lsim 30$\%.
%---------------------------------------------------

\section{The two-component structure function of the
pomeron and the factorization breaking}

%-------------------------
%                           Section 4
%-------
The two fluxes $\phi_{\Pom}(x_{\Pom})$
and $f_{\Pom}(x_{\Pom})$ are not identical (Fig.~2),
because their $x_{\Pom}$-dependence is dominated by
$\sigma(x_{\Pom},r\sim R_{val})$ and
$\sigma(x_{\Pom},r\sim R_{sea})$, respectively, the latter having
a faster growth with ${1\over x_{\Pom}}$ [11,15]. This is a solid
dynamical prediction from the subasymptotic BFKL pomeron, to be
contrasted to conjectured forms of the universal flux of pomerons
[6,7]. Only at very
large ${1\over x_{\Pom}} \gg 1$, well beyond the kinematical range of
HERA, the two fluxes shall
 have a similar ${1\over x_{\Pom}}$-dependence.
Then, the two-component convolution formula for the diffractive DIS
cross section reads:
\arr
x_{\Pom}\left.{d\sigma_{D}
 \over dt\,dx_{\Pom} }\right|_{t=0} =
{4\pi^{2} \alpha_{em}
\over Q^{2}} \cdot
{  \sigma_{tot}(pp) \over 16\pi}\cdot
\left[
\phi_{\Pom}(x_{\Pom})
F_{val}^{\Pom}(\beta,Q^{2})+
f_{\Pom}(x_{\Pom})
F_{sea}^{\Pom}(\beta,Q^{2})\right] \nonumber\\
={4\pi^{2} \alpha_{em}
\over Q^{2}} \cdot
{  \sigma_{tot}(pp) \over 16\pi}\cdot
\Phi_{D}(x,\beta,Q^{2},t=0)
\,\, .
\label{eq:4.1}
\endarr
It is convenient to study the factorization breaking in terms of
the two ratios
\beq
r_{1}(\beta,x_{\Pom}) = {\Phi_{D}(x=x_{\Pom}\beta,\beta,Q^{2},0) \over
\Phi_{D}(x=x_{\Pom}^{0}\beta,\beta,Q^{2},0)}
{}~~~ {\rm and }~~~
r_{2}(x_{\Pom},\beta)=
{\Phi_{D}(x=x_{\Pom}\beta,\beta,Q^{2},0) \over
\Phi_{D}(x=x_{\Pom}\beta_{0},\beta_{0},Q^{2},0)}
\label{eq:4.3}
\endeq
and study their $x_{\Pom}$ and $\beta$ dependence, varying $x$ at fixed
$\beta$ and $x_{\Pom}$, respectively. If the two fluxes were identical,
$\phi_{\Pom}(x_{\Pom})=
f_{\Pom}(x_{\Pom})$, then Eq.~(\ref{eq:4.1}) would have
reduced to the naive parton model convolution (\ref{eq:1.5})
with the consequence that $r_{1}(\beta,x_{\Pom}) =
f_{\Pom}(x_{\Pom})$ would be
independent of $\beta$, whereas $r_{2}(x_{\Pom},\beta)$
would be independent
of $x_{\Pom}$. Our two-component
picture predicts a strong factorization
breaking,
\beq
r_{1}(\beta \gsim 0.3,x_{\Pom}) \approx \phi_{\Pom}(x_{\Pom}) \neq
r_{1}(\beta \ll 0.1,x_{\Pom}) \approx f_{\Pom}(x_{\Pom}) \, ,
\label{eq:4.4}
\endeq
with the $\sim 80$\% ($\sim 30$\%)
 departure of $f_{\Pom}(x_{\Pom})$ from
$\phi_{\Pom}(x_{\Pom})$ as $x_{\Pom}$
decreases from $x_{\Pom}=0.03$ down to $x_{\Pom}=10^{-4}~(x_{\Pom}=
10^{-3})$,
see Fig.~2. Similarly, the results from our two-component
picture for $r_{2}(x_{\Pom}=0.03,\beta)$ and
$r_{2}(x_{\Pom}=0.0001,\beta)$ show
a large, $\sim 50\%$, factorization breaking (Fig.~4) at $\beta <0.1$.

Above we focused on the forward diffraction dissociation,
$t=0$. In [1-3,16] we argued that excitation of the valence of
the pomeron is the counterpart of diffraction production of
resonances in the hadronic scattering and/or real photoproduction,
which have the slope of the $t$-dependence close to that of the
diffraction slope of elastic scattering $B_{el}$, whereas
excitation of the sea of the pomeron is the counterpart
of the triple-pomeron regime with $B_{3\Pom} \sim {1\over 2}B_{el}$.
The hadronic and real photoproduction data give $B_{3\Pom} \approx
6$\,GeV$^{-2}$ [19,29] to an
 uncertainty $\lsim 25$\%.
Then, the extension of (\ref{eq:4.1}) to
the non-forward diffractive DIS, which can be studied at HERA when
the ZEUS and H1 leading proton spectrometers will be in operation,
reads
\arr
\Phi_{D}(x,\beta,Q^{2},t)=
\phi_{\Pom}(x_{\Pom})
F_{val}^{\Pom}(\beta,Q^{2})\exp(-B_{el}|t|)+
f_{\Pom}(x_{\Pom})
F_{sea}^{\Pom}(\beta,Q^{2})\exp(-B_{3\Pom}|t|) \, .
\label{eq:4.5}
\endarr
Because of the different $t$ dependence of
$\phi_{\Pom}(x_{\Pom})\exp(-B_{el}|t|)$
and $f_{\Pom}(x_{\Pom})\exp(-B_{3\Pom}|t|)$
we predict a $t$-dependent
factorization breaking. Finally,
the $t$-integrated mass spectrum equals
\arr
x_{\Pom}{d\sigma_{D}  \over dx_{\Pom} } =
{4\pi^{2} \alpha_{em}\over Q^{2}} \cdot
{  \sigma_{tot}(pp) \over 16\pi B_{3\Pom}}\cdot
\left[
{B_{3\Pom} \over B_{el}}\cdot
\phi_{\Pom}(x_{\Pom})
F_{val}^{\Pom}(\beta,Q^{2})+
f_{\Pom}(x_{\Pom})
F_{sea}^{\Pom}(\beta,Q^{2})\right]
\,\,
\label{eq:4.6}
\endarr
and is different [1-3]
from the mass spectrum in the forward, $t=0$,
dissociation for the emergence of the factor
$B_{3\Pom}/ B_{el}\approx {1/2}$
in the first term in the {\sl r.h.s.} of (\ref{eq:4.6}), which
makes the
relative contribution from $M^{2} \sim Q^{2}$ to the inclusive
mass spectrum smaller than at $t=0$. This prediction can be
tested at HERA after the data taking with leading proton
spectrometer.
%------------------------------------------------------
\section{Diffractive contribution to $F_{2}^{p}(x,Q^{2})$}
%---------------------------------
%                              Section 5
%---------------------------
The total diffraction dissociation cross section $\sigma_{D}
= \int dt dM^{2} [d\sigma_{D}/dt dM^{2}]$ defines the diffractive
contribution
$F_{2}^{D}(x,Q^{2})=
{Q^{2} \over 4\pi^{2}\alpha_{em}}\sigma_{D}$
to the proton structure function $F_{2}^{p}(x,Q^{2})$,
\arr
F_{2}^{D}(x,Q^{2})=
{ \sigma_{tot}(pp) \over 16\pi B_{3\Pom}}
\int_{x}^{x_{\Pom}^{c}}{dx_{\Pom} \over x_{\Pom}}
\left[{B_{3\Pom}\over B_{el}}\cdot
\phi_{\Pom}(x_{\Pom})F_{val}^{\Pom}({x\over x_{\Pom}},Q^{2})+
f_{\Pom}(x_{\Pom})F_{sea}^{\Pom}({x\over x_{\Pom}},Q^{2})\right]
\,,
\label{eq:5.1}
\endarr
where the numerical factor $\sigma_{tot}(pp)/16\pi B_{3\Pom}
\approx 0.3$ to an,
$\lsim 25$\%, uncertainty coming from the
uncertainty in $B_{3\Pom}$, which can eventually be reduced
with the advent of the HERA measurements of $B_{3\Pom}$.
Here $x_{\Pom}^{c}$ is subject to the experimental
(pseudo)rapidity gap cutoff used to define the diffractive DIS,
$\Delta \eta \gsim \Delta\eta_{c}\approx \log{1\over x_{\Pom}^{c}}$.
In hadronic interactions with the recoil-proton tagging of
diffraction dissociation, the pomeron exchange mechanism was
shown to dominate at $x_{\Pom} \lsim x_{\Pom}^{c}=
$ (0.05-0.1) [19,29]. The
preliminary data from HERA correspond to
rather a conservative
cutoff $x_{\Pom}^{c} \lsim 0.01$ [30,31].

In Fig.~5 we present our predictions for the diffractive
structure function $F_{2}^{D}(x,Q^{2})$. The strikingly
weak $Q^{2}$-dependence of $F_{2}^{D}(x,Q^{2})$ has its
origin [1,16,32] in the fact that
$F_{val}^{\Pom}(\beta,Q^{2})$ and $\,F_{sea}^{\Pom}(\beta,Q^{2})$
enter the integrand of (\ref{eq:5.1}) at large values of $\beta =
x/x_{\Pom}$ such that the predicted scaling violations, shown in Fig.~3,
are still weak. Furthermore, the fluxes
$\phi_{\Pom}(x_{\Pom})$ and $f_{\Pom}(x_{\Pom})$ rise
towards small $x_{\Pom}$,
enhancing the contribution from large $\beta$ and further
minimizing the $Q^{2}$ dependence of $F_{2}^{D}(x,Q^{2})$. We
predict a steep rise of the diffractive structure function
at large ${1\over x}$, which predominantly comes from the rapid
rise of the flux function $f_{\Pom}(x_{\Pom})$.
Fig.~5 also shows
a sensitivity of predicted $F_{2}^{D}(x,Q^{2})$ to the
value of $x_{\Pom}^{c}$ (the minimal rapidity gap $\Delta \eta_{c}$
). We find a good
agreement with the H1 estimates [31] for
$F_{2}^{D}(x,Q^{2})$. Notice that our
calculation does not include a possible enhancement of the
H1 and ZEUS values of $F_{2}^{D}(x,Q^{2})$ for the unrejected
diffraction excitation of protons into proton resonances and/or
multiparticle states which escaped into the beam pipe. From
the hadronic interaction data [19], we can conclude that
possible overestimation of $F_{2}^{D}(x,Q^{2})$ by the H1 and ZEUS
can not exceed, and, presumably, is significantly smaller than,
$30 \%$.
Our results for the ratio
$
r_{D}(x,Q^{2})={F_{2}^{D}(x,Q^{2})/ F_{2}^{p}(x,Q^{2})}
$
are shown in Fig.~6.
The steady
decrease of $r_{D}(x,Q^{2})$ with $Q^{2}$ was predicted
in [1,16] and predominantly comes from the scaling violations
in the proton structure function. The overall agreement with
the H1 [31] and ZEUS [33] results is good.
%-----------------------------------------------------

\section{Conclusions}

%-------------------------
The purpose of this study has been a calculation of the parton
distributions in the pomeron starting with the microscopic
dipole BFKL pomeron. We have shown that the pomeron must be
endowed with a two-component structure function, the two
components being related to the initial valence $q\bar{q}$ and
the valence $gg$ states of the pomeron and entering
the description of diffractive DIS with different fluxes
of pomerons in the proton. The predicted breaking of the
conventional parton-model factorization is strong and
can be tested with higher precision data from HERA.
We have presented parameter-free predictions for the
pomeron structure function and for diffractive contribution
to the proton structure function, which agree with the first
experimental data from HERA.

It is worthwhile to notice that
the two-component structure function of the pomeron is by itself
an approximation. For instance, the $x_{\Pom}$- and $r$-dependence of
the dipole cross section
$\sigma(x_{\Pom},r)$ do not factorize [2-5,11,15]
and the dipole size
$r\sim R_{val}(x_{\Pom}),R_{sea}(x_{\Pom})$, from which comes the
dominant contribution in (\ref{eq:2.1}) and (\ref{eq:3.2}),
changes with $x_{\Pom}$. Consequently, the span of the QCD evolution
which is given [11,21] by $\log[Q^{2}R_{val}^{2}(x_{\Pom})], \,
\log[Q^{2}R_{sea}^{2}(x_{\Pom})]$, it changes with $x_{\Pom}$,
breaking
the factorization (\ref{eq:2.5},\ref{eq:4.1}) slightly. Numerically, in
the region of $(x_{\Pom},Q^{2})$
of interest at HERA, the variations
of $R_{val}(x_{\Pom})$ and $R_{sea}(x_{\Pom})$
are still much smaller than
the large difference between $R_{val}$ and $R_{sea}$ which is
the origin of the two-component description. Because of the
small $R_{sea}$, at $Q^{2} \lsim Q_{\Pom}^{2}=10$\,GeV$^{2}$,
significant departure of the $Q^{2}$ evolution of $F_{sea}^{\Pom}
(\beta,Q^{2})$ from the GLDAP evolution is possible. A detailed
description of transition from the real photoproduction $Q^{2}=0$
to DIS will be presented elsewhere.
\medskip\\
{\bf Acknowledgements:} B.G.Zakharov thanks J.Speth for the
hospitality at the Institut f\"ur Kernphysik, KFA, J\"ulich.
This work was partially supported by the INTAS grant 93-239.

\pagebreak

\pagebreak
{\bf \Large Figure captions}
\begin{itemize}
\item[Fig.1]
 - The diffraction excitation diagrams describing DIS on (a) valence
$q\bar{q}$ of the pomeron,
(c) the valence-glue generated
sea of the pomeron and (b,c) the $Q^{2}$-evolution effects.

\item[Fig.2]
 - Predictions from the dipole BFKL pomeron for flux functions
$\phi_{\Pom}(x_{\Pom})$ (dashed curve) and $f_{\Pom}(x_{\Pom})$
(solid curve) for the
$F_{val}^{\Pom}(\beta,Q^{2})$ and $F_{sea}^{\Pom}(\beta,Q^{2})$
components of the pomeron structure function, respectively,
and the ratio of the two fluxes (the bottom box).

\item[Fig.3]
- Predictions from the dipole BFKL pomeron for the
$Q^{2}$-evolution
of components $F_{val}^{\Pom}(\beta,Q^{2})$ (solid
curves)
and $F_{sea}^{\Pom}(\beta,Q^{2})$ (dashed curves)
of the pomeron structure function.

\item[Fig.4]
- Predicted factorization breaking in the $\beta$ distribution
$r_{2}(x_{\Pom},\beta)$ at $x_{\Pom}=0.03$ (solid curve)
and $x_{\Pom}=0.0001$ (dashed curve). We take $\beta_{0}=0.5$
and $Q^{2}=25$GeV$^{2}$.

\item[Fig.5]
- Predictions from the dipole BFKL pomeron for the diffractive
contribution $F_{2}^{D}(x,Q^{2})$ to the proton structure function.
The solid and dotted curves are for $x_{\Pom}^{c}=0.01$ and
$Q^{2}=10$ and $Q^{2}=100$\,GeV$^{2}$, respectively. The dashed
and dot-dashed curves are for $Q^{2}=10$\,GeV$^{2}$ and
rapidity-gap cuts $x_{\Pom}^{c}=0.003$ and $x_{\Pom}^{c}=0.03$,
respectively. The data points are from the H1 experiment [31].
\item[Fig.6]
- Predictions for the ({\bf a,b,c}) $Q^{2}$ and ({\bf d})
$x$ dependence of
the fraction $r_{D}(x,Q^{2})=F_{2}^{D}(x,Q^{2})/
F_{2}^{p}(x,Q^{2})$ of DIS on protons which goes via diffraction
dissociation of photons for the cut $x_{\Pom}^{c}=0.01$. The data
points are from the H1 [31] and ZEUS
[33] experiments. The data points shown in the box ({\bf d}) are
for the lowest $Q^{2}$ bins in boxes ({\bf a-c}).

\end{itemize}

\begin{thebibliography}{299}

\bibitem{1} %%%
N.N.~Nikolaev and B.G.~Zakharov, {\it Z. Phys.} {\bf C53} (1992) 331.

\bibitem{2} %%%
N.N.~Nikolaev and B.G.~Zakharov,
The triple-pomeron regime and the structure function of the
in the diffractive deep inelastic scattering at very small
$x$, {\sl J\"ulich preprint} {\bf KFA-IKP(Th)-1993-17}, June 1993,
to appear in
{\sl Z. f. Physik } {\bf Cxx} (1994) xxx.

\bibitem{3} %%%
N.N.~Nikolaev and B.G.~Zakharov,
{\sl JETP} {\bf 78} (1994) 598.

\bibitem{4} %%%
N.N.Nikolaev, B.G.Zakharov and V.R.Zoller,
{\sl JETP Letters}  {\bf 59} (1994) 8.

\bibitem{5} %%%
N.N.Nikolaev, B.G.Zakharov and V.R.Zoller,
{\sl Phys. Lett.} {\bf B328} (1994) 486;
{\sl Zh. Exp. Teor. Fiz.} {\bf 105} (1994) 1498-1524.

%===============          ===   5    =============

\bibitem{6} %%%
K.A.Ter-Martirosyan, {\sl Phys.Lett.} {\bf B44} (1973) 179;
A.B.Kaidalov and K.A.Ter-Martirosyan, {\sl Nucl.Phys.}
{\bf B75} (1974) 471.

\bibitem{7} %%%
G.Ingelman and P.Schlein, {\sl Phys. Lett.} {\bf B152} (1985) 256;
H.Fritzsch and K.H.Streng, {\sl Phys.Lett.} {\bf B164} (1985) 391;
E.L.Berger, J.C.Collins, D.E.Soper and G.Sterman,
{\sl Nucl.Phys.} {\bf B286} (1987) 704.

\bibitem{8} %%%
A.Donnachie and P.V.Landshoff, {\sl Phys.Lett.} {\bf B191}
(1987) 309; {\sl Nucl.Phys.} {\bf B303} (1988) 634.

\bibitem{9} %%
E.A.Kuraev, L.N.Lipatov and V.S.Fadin, {\sl Sov.Phys. JETP}
{\bf 44} (1976) 443; {\bf 45} (1977) 199;
Ya.Ya.Balitsky and L.N.Lipatov, {\sl Sov. J. Nucl. Phys.}
{\bf 28} (1978) 822;
L.N.Lipatov, {\sl Sov. Phys. JETP} {\bf 63} (1986) 904;
L.N.Lipatov. Pomeron in Quantum Chromodynamics. In: {\sl Perturbative
Quantum Chromodynamics}, editor A.H.Mueller, World Scientific, 1989.

\bibitem{10} %%%
A.Mueller and B.Patel, {\it Nucl. Phys.} {\bf B 425} (1994) 471.
%==============           ====    10   ========

\bibitem{11} %%%
N.N.~Nikolaev and B.G.~Zakharov,
{\sl Phys. Lett.} {\bf B327} (1994) 149;
{\bf B327} (1994) 157.

\bibitem{12} %%%
N.N.~Nikolaev and B.G.~Zakharov,
{\sl Phys. Lett.} {\bf B332} (1994) 177.

\bibitem{13} %%%
V.N.~Gribov and L.N.~Lipatov, {\it Sov. J. Nucl. Phys.} {\bf 15} (1972)
438; L.N.~Lipatov, {\it Sov. J. Nucl. Phys.} {\bf 20} (1974) 181;
Yu.L.~Dokshitser, {\it Sov. Phys. JETP} {\bf 46} (1977) 641;
G.~Altarelli and G.~Parisi, {\it Nucl. Phys.} {\bf B126} (1977) 298.

\bibitem{14} %%%%
M.Genovese, N.N.Nikolaev and B.G.Zakharov,
Direct calculation of the triple-pomeron coupling in DIS
and real photoproduction, {\sl J\"ulich preprint}
{\bf KFA-IKP(TH)-1994-36} and Torino Univ. preprint {\bf DFTT
41/94}, October 1994, submitted to
{\sl Phys. Lett.} {\bf B}.

\bibitem{15} %%%%
J.Nemchik, N.N.Nikolaev and B.G.Zakharov,
Scanning the BFKL pomeron in elastic production of vector mesons
at HERA, {\sl J\"ulich preprint}
{\bf KFA-IKP(TH)-1994-17}, May 1994, to appear in
{\sl Phys. Lett.} {\bf Bxxx} (1994) xxx.

%==============           ====    15   ========

\bibitem{16} %%%%
N.N.~Nikolaev and B.G.~Zakharov, {\it Z. Phys.} {\bf C49} (1991) 607.

\bibitem{17} %%%%
ZEUS Collaboration: M.Derrick et al., {\sl Phys. Lett.} {\bf B316}
(1993) 412; H1 Collaboration: I.Abt et al., {\sl Nucl. Phys.}
{\bf B407} (1993) 515.

\bibitem{18} %%%
V.Barone, M.Genovese, N.N.Nikolaev, E.Predazzi and B.G.Zakharov,
{\sl Z. Phys.} {\bf C58} (1993) 541.

\bibitem{19} %%%%
For reviews, see e.g.:
A.B.Kaidalov, {\sl Phys. Rep.} {\bf 50} (1979) 157;
G.Alberi and G.Goggi, {\sl Phys. Rep.} {\bf 74} (1981) 1;
K.Goulianos, {\sl Phys. Rep.} {\bf 101} (1983) 169.

\bibitem{20} %%%
E.V.Shuryak, {\sl Rev. Mod. Phys.} {\bf 65} (1993) 1, and
references therein.

%==============           ====    20   ========

\bibitem{21} %%%%
N.N.Nikolaev and B.G.Zakharov,
{\sl Phys. Lett.} {\bf B332} (1994) 184.

\bibitem{22} %%%%
P.L.Frabetti et al., {\sl Phys. Lett.} {\bf B316} (1993) 197.

\bibitem{23} %%%%
NMC collaboration: M.Arneodo et al., {\bf CERN-PPE/94-146}.

\bibitem{24} %%%%
NMC Collaboration: M.Arneodo et al., {\sl Phys. Lett.} {\bf B332}
(1994) 195.

\bibitem{25} %%%%
E665 Collaboration: M.R.Adams et al., {\bf FERMILAB-PUB-94/233-E}.

%==================       =====    25   =========

\bibitem{26} %%%
B.Z.Kopeliovich and B.G.Zakharov, {\sl Phys. Rev.} {\bf D44} (1991)
3466; O.Benhar, B.Z.Kopeliovich, Ch.Mariotti, N.N.Nikolaev and
B.G.Zakharov, {\sl Phys. Rev. Lett.} {\bf 69} (1992) 1156.

\bibitem{27} %%%
B.Z.Kopeliovich, J.Nemchik, N.N.Nikolaev and B.G.Zakharov,
{\sl Phys. Lett.} {\bf B324} (1994) 469.

\bibitem{28} %%%%
J.Nemchik, N.N.Nikolaev and B.G.Zakharov,
Color transparency after the NE18 and E665 experiments: Outlook
and perspectives at CEBAF, {\sl J\"ulich preprint}
{\bf KFA-IKP(TH)-1994-20}, June 1994, to be published in
Proceedings
of the Workshop on CEBAF at Higher Energies, 14-16 March 1994,
CEBAF.

\bibitem{29} %%%%
T.J.Chapin et al., {\sl Phys. Rev. } {\bf D31} (1985) 17.

\bibitem{30} %%%%
ZEUS Collaboration: M.Derrick et al., {\sl Phys. Lett.} {\bf B315}
(1993) 481.

%----------------------------------  30  -------------

\bibitem{31} %%%%
H1 Collaboration: T.Ahmed al., {\bf DESY 94-133}; S.Levonian,
Mini-School on Diffraction at HERA, DESY, May 4-7, 1994.


\bibitem{32} %%%%
N.N.Nikolaev, EMC effect and quark degrees of freedom in nuclei:
facts and fancy. {\sl Oxford Univ. preprint} {\bf OU-TP 58/84}
(1984); Also in: {\sl Multiquark Interactions and Quantum
Chromodynamics.} Proc. VII Intern. Seminar on Problems of High
Energy Physics, 19-26 June 1984, Dubna, USSR.

\bibitem{33} %%%%
ZEUS Collaboration: R.Devenish, Mini-School on Diffraction at HERA,
DESY, May 4-7, 1994.

\end{thebibliography}
\end{document}